\documentclass[pra,twocolumn,superscriptaddress,amsmath,amssymb,showpacs]{revtex4}
\pdfoutput=1
\usepackage{amssymb}
\usepackage{graphicx}% Include figure files
\usepackage{dcolumn}% Align table columns on decimal point
\usepackage{bm}% bold math

\begin{document}

\title{Filamentation of ultrashort laser pulses in silica glass and KDP crystal: A comparative study}
\author{J\'er\'emie Rolle}
\affiliation{CEA, DAM, DIF, 91297 Arpajon, France}
\author{Luc Berg\'e}
\affiliation{CEA, DAM, DIF, 91297 Arpajon, France}
\author{Guillaume Duchateau}
\affiliation{Universit\'e de Bordeaux-CNRS-CEA, Centre Lasers Intenses et Applications, UMR 5107, 351 Cours de la Lib\'eration, 33405 Talence, France}
\author{Stefan Skupin}
\affiliation{Universit\'e de Bordeaux-CNRS-CEA, Centre Lasers Intenses et Applications, UMR 5107, 351 Cours de la Lib\'eration, 33405 Talence, France}

\date{\today}

\begin{abstract}
Ionizing 800-nm femtosecond laser pulses propagating in silica glass and in potassium dihydrogen phosphate (KDP) crystal are investigated by means of a unidirectional pulse propagation code. Filamentation in fused silica is compared with the self-channeling of light in KDP accounting for the presence of defect states and electron-hole dynamics. In KDP, laser pulses  produce intense filaments with higher clamping intensities up to 200~TW/cm$^2$ and longer plasma channels with electron densities above $10^{16}$ cm$^{-3}$. Despite these differences, the propagation dynamics in silica and KDP are almost identical at equivalent ratios of input power over the critical power for self-focusing.
\end{abstract}


\maketitle

\section{Introduction} \label{Sec1}

For two decades, investigations of intense laser pulses interacting with dielectrics have revealed key features in the nonlinear propagation of light and the relatively high robustness of irradiated materials. An intriguing phenomenon is the filamentation of powerful beams in transparent solids, stemming from the interplay between diffraction, chromatic dispersion, Kerr self-focusing and generation of free carriers~\cite{Berge:rpp:70:1633,Tzortzakis:prl:87:213902,Sudrie:prl:89:186601,Berge:pra:81:013817}. From a theoretical point of view, an important issue is the accurate modeling of materials like silica glass (SiO$_2$), potassium dihydrogen phosphate (KH$_2$PO$_4$ or KDP) and its deuterated analog (KD$_2$PO$_4$ or DKDP)~\cite{Hebert:jap:109:123527}. Routinely used in high-power laser systems devoted to, e.g., inertial confinement fusion, these materials are exposed to intense radiation. The precise knowledge of their inherent ionization properties and of the nonlinear light-matter interaction is thus essential to understand how and where laser damage initiated by plasma generation takes place. In particular, the strong increase in laser intensity which occurs during filamentation can be a potential source of damage, because photo-induced ionization produces an electron plasma in the wake of the optical field.

While SiO$_2$ glasses and KDP crystals are mostly exploited in the fusion context for nanosecond long pulses, laser-induced damage has recently been examined for shorter femtosecond pulses, both experimentally and theoretically. One motivation is the understanding of the influence of precursor defects, which may locally generate non-uniform damage zones over nanometer sizes. In KDP, precursors are suspected to be connected with the proton transport in the hydrogen bond network, which induces defects, e.g., oxygen vacancies, leading to hole trapping. For an accurate modeling of the interaction it is important to incorporate the particular band structure of KDP. For instance, it was recently shown that electronic {\it states located in the band gap} (SLG) can serve as intermediate transition states to transfer carriers (electrons) from the valence band to the conduction band, and such transitions involve lower photon numbers than a direct multiphoton process~\cite{Carr:prl:91:127402,Demos:oe:18:13788,Duchateau:prb:83:075114}. Moreover, the trapping rates increase with the laser excitation density, and hole trapping precedes electron trapping~\cite{Martin:prb:55:5799}. In contrast, for silica glass, electron transitions from the valence band to the conduction band are direct and the kinetics of electron trapping (self-trapped excitons) does not depend on the laser intensity.

The goal of this paper is to compare the filamentation of 800-nm femtosecond pulses in silica and KDP. Filamentation is an omnipresent phenomenon in high-intensity pulse propagation and hence of utmost importance for the understanding of possible laser induced damage. In particular, the maximum intensity occuring inside a material is given by the so-called clamping intensity, determined by the self-channeling dynamics. Unfortunately, the clamping intensity in solids is not accessible by direct experimental measurements. Therefore, estimates extracted from numerical simulations usually provide valuable information. This clamping intensity is an important parameter for the interaction of laser pulses with crystal defects~\cite{Sasaki:jcg:99:820} and possible dopants~\cite{Meena:43:166:crt}.

The quantitative reliability of simulation results crucially depends on the material models, in particular the accurate description of the free carrier densities.
Our selected ionization models consider either direct transitions from the valence band to the conduction band only (silica) or additional transitions due to the presence of SLGs (KDP). Different photo-ionization descriptions are used, from the complete Keldysh rate with tunnel ionization~\cite{Keldysh:spjetp:20:1307} to purely multiphoton ionization (MPI) rates. The growth of the electron density in the conduction band is limited by electron relaxation, whose associated time scale depends on the trapped-hole population and their ability to trap free electrons. 

The paper is organized as follows. Section~\ref{Sec2} recalls the propagation model and ionization schemes, with emphasis on KDP crystals. Section~\ref{Sec3} compares the filamentation dynamics of short 50-fs pulses in silica glass and in KDP. Despite the comparable size of band gaps in silica and KDP, we report significant differences in the ionization rates leading to three times higher clamping intensities in KDP and a longer plasma channel. The former effect is caused by a weak 5-photon ionization cross-section; the latter is supported by transitions from the SLG states. At the same time, we identify very similar spatio-temporal filamentation dynamics in both silica and KDP after rescaling the pulse power. Section~\ref{Sec4} examines longer pulses, for which we expect to enhance the impact of defect states over longer relaxation times, since the associated electron recombination time lies in the picosecond timescale. Most of the filamentation characteristics for short pulses are, however, retrieved.

\section{Model equations} \label{Sec2}

We integrate the following axial-symmetric pulse propagation model for filamentation~\cite{Berge:rpp:70:1633}
\begin{equation}
\begin{split}
\partial_z \mathcal{E} & = \frac{i}{2k_0}\hat{T}^{-1}\frac{1}{r}\partial_r r \partial_r
\mathcal{E} + i\hat{\mathcal{D}}\mathcal{E} 
-i\frac{k_0}{2n_0^2\rho_c}\hat{T}^{-1}\rho\mathcal{E} \\
 & -\frac{\sigma}{2}\rho\mathcal{E} + i\frac{\omega_0}{c} n_2 \hat{T} \int\limits_{-\infty}^t \mathcal{R}(t-t') \left|\mathcal{E}(t') \right|^2 dt' \mathcal{E} -\frac{1}{2}\beta\mathcal{E},
 \label{krausz_radial}
\end{split}
\end{equation}
where $z$ is the propagation variable, $t$ is the retarded time in a frame moving with group velocity $1/k^{(1)}$ at center frequency $\omega_0$, 
$k(\omega)=n(\omega)\omega/c$, $n_0=n(\omega_0)$, $\hat{T}=1+(i/\omega_0)\partial_t$, and 
$\hat{\mathcal{D}}=\sum_{m\geq2} k^{(m)}\left(i\partial_t \right)^m/m!$ is the dispersion
operator formally involving the derivatives $k^{(m)}=\left. \partial^m k/\partial \omega^m \right|_{\omega_0}$.
The electric field envelope $\mathcal{E}$ is normalized such that $I=|\mathcal{E}|^2$ equals the laser intensity~\footnote{Note that due to large phase mismatch we do not need to account for generation of harmonic frequencies, which allows us to use an envelope description for the optical field.}. $\mathcal{R}(t)$ denotes the time-response function of the Kerr nonlinearity, $\rho_c = 1.73 \times 10^{21}$~cm$^{-3}$ is the critical plasma density at 800~nm, and $\rho$ is the electron density produced in the conduction band.

\subsection{Fused silica}
Our modelling of the optical properties of fused silica follows Ref.~\cite{Agrawal:NFO:01}. Linear dispersion is included via the Sellmeier formula
\begin{equation}
n^2(\lambda[\mu \mbox{m}]) = 1+\sum\limits_{j=1}^3 \frac{B_j \lambda^2}{\lambda^2-\lambda_j^2},
\label{eq:sellsi}
\end{equation}
whose parameters are summarized in Table~\ref{table:sellmeier}. We can evaluate the linear refraction index $n_0 = 1.454$ and the group velocity dispersion (GVD) coefficient $k^{(2)} \simeq 362$ fs$^2$/cm at wavelength $\lambda=800$ nm. The response function
\begin{equation}
 \mathcal{R}(t) =(1-f)\delta(t)+f\theta(t)\frac{1+\omega_R^2 \tau_R^2}{\omega_R\tau_R^2}\mbox{e}^{-\frac{t}{\tau_R}}\sin(\omega_R t),
\end{equation}
contains a Raman-delayed contribution with ratio $f=0.18$, rotational delay time $\tau_K = 32$ fs and resonance frequency $\omega_R = 0.082$ fs$^{-1}$. The nonlinear index for glass is $n_2=3.2\times 10^{-16}$~cm$^2$/W \cite{Agrawal:NFO:01,Skupin:pd:220:14}, so that the critical power for self-focusing, $P_{\rm cr} \simeq \lambda_0^2/(2\pi n_0 n_2)$, takes the value $P_{\rm cr}^{\rm silica} = 2.19$~MW.

\begin{table}[ht]
\begin{tabular}{|r|r||r|r|}
\hline $B_1$ & 0.6961663 & $\lambda_1$ & 0.0684043 \tabularnewline
\hline $B_2$ & 0.4079426 & $\lambda_2$ & 0.1162414 \tabularnewline
\hline $B_3$ & 0.8974794 & $\lambda_3$ & 9.896161 \tabularnewline
\hline
\end{tabular}
\caption{Parameters for silica dispersion Eq.~(\ref{eq:sellsi}).}\label{table:sellmeier}
\end{table} 

The electron density in the conduction band is governed by 
\begin{equation}
\label{silica}
\partial_t \rho  = W(I)\rho_{\rm{nt}}+\sigma \rho  I/U_i-\rho /\tau_{\rm{rec}},
\end{equation}
where the ionization rate $W(I)$ is the complete Keldysh rate \cite{Keldysh:spjetp:20:1307}, considering the gap potential $U_i = 9$~eV \cite{Audebert:prl:73:1990,French:ssom:28:63}. The electron collision time $\tau_c=20$~fs determines the inverse Bremsstrahlung cross-section $\sigma = 6.57 \times 10^{-19}$~cm$^2$ \cite{Sudrie:prl:89:186601}, and the electron recombination time can be estimated as $\tau_{\rm{rec}} = 150$~fs \cite{Tzortzakis:prl:87:213902,Audebert:prl:73:1990}. As long as the density $\rho$ remains small compared to the neutral density $\rho_{\rm{nt}} = 2.1 \times 10^{22}$~cm$^{-3}$, the model gives reliable results. Related ionization losses resulting from photo-ionization can be included in the propagation equation~(\ref{krausz_radial}) via $\beta(I)  = U_iW(I)\rho_{\rm{nt}}/I$.

\subsection{KDP crystal}

For birefringent KDP crystals, we assume the laser beam in ordinary polarization and neglect second harmonic generation due to the large phase mismatch. We use the linear dispersion law given in Ref.~\cite{Bass:HOO:2009} 
\begin{equation}
\label{dispkdp}
n^2(\lambda[\mu \mbox{m}]) = 2.259276 + \frac{0.01008956}{\lambda^2 - 0.0129426} + \frac{13.00522 \lambda^2}{\lambda^2 - 400},
\end{equation}
yielding the linear refraction index $n_0 = 1.502$ and GVD coefficient $k^{(2)} = 274$ fs$^2$/cm. The Kerr response contains only an instantaneous 
contribution, $\mathcal{R}(t)=\delta(t)$, and the nonlinear coefficient $n_2=1.56\times 10^{-16}$~cm$^2$/W~\cite{Duchateau:prb:83:075114} leads to the critical power for self-focusing $P_{\rm cr}^{\rm KDP} = 4.35$~MW.

\begin{figure}
\includegraphics[width=0.8\columnwidth]{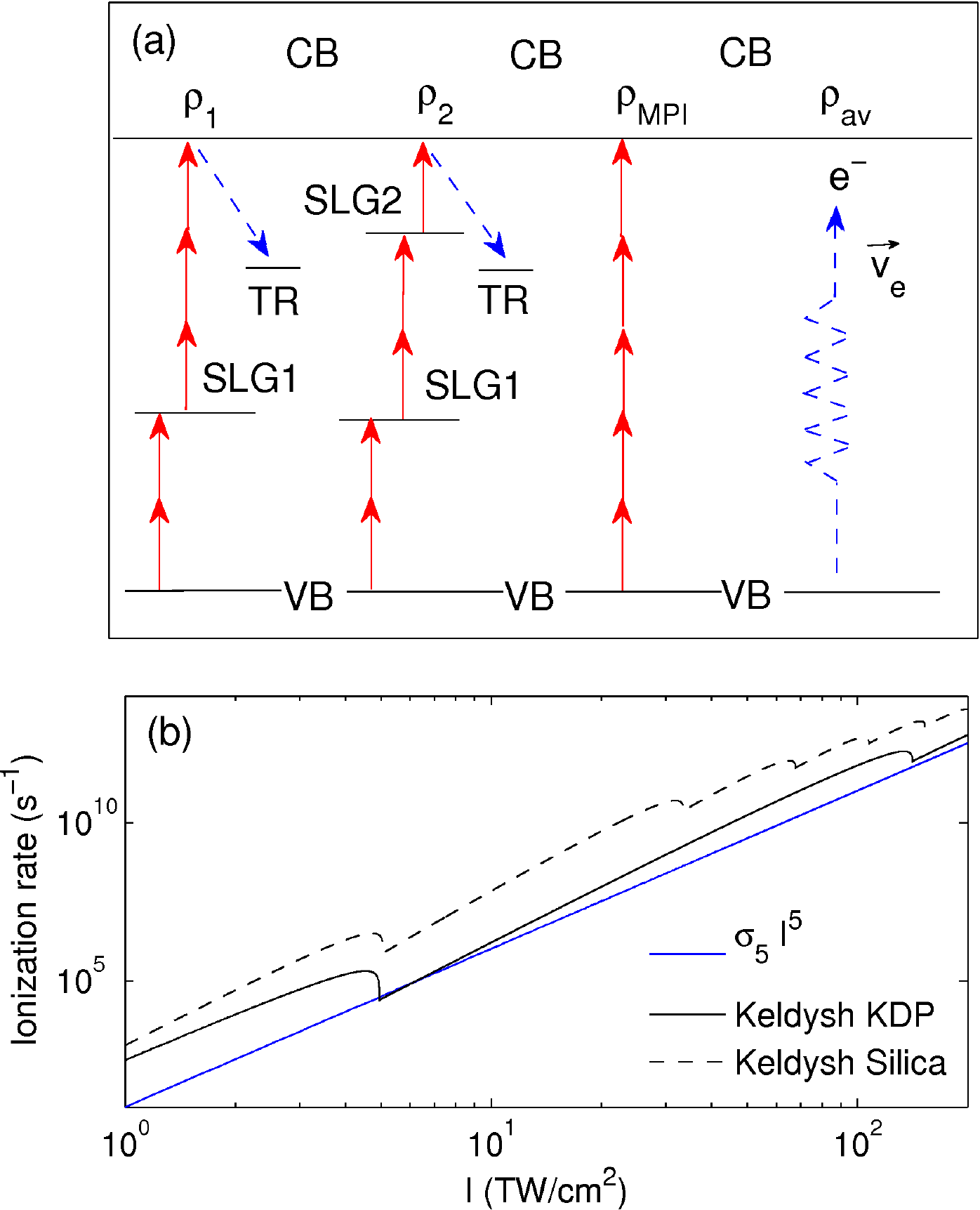}
\caption{(a) Schematic illustration of the four ionization channels Eqs.~(\ref{eq:ionization_kdp}) considered for the KDP crystal: 3-photon ionization from a defect state SLG1, 1-photon transition from a defect state SLG2 close to the conduction band,  direct 5-photon ionization from the valence band, and avalanche ionization. (b) Comparison of ionization rates for multiphoton transitions at 800 nm. The black curve refers to the Keldysh rate \cite{Keldysh:spjetp:20:1307} for silica (dashed curve) and for KDP (solid curve). The blue (dark gray) solid line shows the rate $\sigma_5 I^5$ used for KDP in Eq.~(\ref{eq:mpi_kdp}).} 
\label{Fig1}
\end{figure}

As far as the electron density in the conduction band of KDP is concerned, we resort to a model taking into account various defect states (SLGs) in the energy gap $U_i=7.7$~eV between valence band and conduction band. This model was recently developed in~\cite{Duchateau:prb:83:075114} and validated against femtosecond pump probe experiments. Here, four different ionization channels contribute to the electron population in the conduction band, namely, 3-photon ionization from a defect state SLG1 located at $\sim$ 3.1~eV above the valence band, 1-photon transition from a defect state SLG2 close to the conduction band,  direct 5-photon ionization from the valence band, and impact ionization. All four mechanisms are illustrated schematically in Fig.~\ref{Fig1}(a). Each ionization channel is treated independently from the others, as well as their respective recombination mechanisms.

Assuming that defect state SLG1 is initially filled and its small electronic density $\rho_{\rm{SLG1}} \ll \rho_{\rm{nt}}=2.1 \times 10^{22}$~cm$^{-3}$ does not change during the interaction, the equation for the 3-photon ionization channel reads
\begin{subequations}
\label{eq:ionization_kdp}
\begin{equation}
\partial_t \rho_{1}=\sigma_3 I^3 \rho_{\rm{SLG1}} - \rho_{1}/\tau_{1},
\end{equation}
where $\rho_1$ is the carrier density delivered in the conduction band and $\tau_{1}$ is the recombination time provided in Table II.

The system of equations describing 1-photon ionization from defect state SLG2 is a bit more involved \cite{Duchateau:prb:83:075114}. It takes into account the electron-hole dynamics and the population of SLG2 from SLG1 via a 2-photon process:
\begin{align}
\begin{split}
\partial_t \rho_{2} & = \sigma_1 I \rho_{\rm{SLG2}}  - \sigma_cv \rho_{2} \left( \rho_{\rm{th}} - \rho_{\rm{tr}} \right) 
\end{split} \label{eq:rho2_kdp}\\
\partial_t \rho_{\rm{tr}} & = \sigma_cv
\rho_{2} \left( \rho_{\rm{th}} - \rho_{\rm{tr}} \right) \label{eq:rhotr_kdp}\\\
\partial_t \rho_{\rm{fh}} & = \sigma_1 I \rho_{\rm{SLG2}} - \rho_{\rm{fh}}/\tau_{\rm{fh}}\\
\partial_t \rho_{\rm{th}} & = \rho_{\rm{fh}}/\tau_{\rm{fh}}\\
\partial_t \rho_{\rm{SLG2}} & = \sigma_2 I^2 \rho_{\rm{SLG1}}-\sigma_1 I \rho_{\rm{SLG2}},
\end{align}
where $\rho_{2}$ and $\rho_{\rm{tr}}$ are the densities of free and trapped electrons, respectively; $\rho_{\rm{fh}}$ and $\rho_{\rm{th}}$ are the densities of free and trapped holes. The term $\sigma_cv$ accounting for electron capture is discussed hereafter.

Originally discarded in Ref.~\cite{Duchateau:prb:83:075114} for more moderate intensities, direct 5-photon ionization employs an MPI rate and the related density of free electrons is governed by 
\begin{equation}
\partial_t \rho_{\rm{MPI}}=\sigma_5 I^5 \rho_{\rm{nt}} - \rho_{\rm{MPI}}/\tau_{\rm{rec}},
\label{eq:mpi_kdp}
\end{equation}
where $\sigma_5$ is the 5-photon ionization cross section and $\tau_{\rm{rec}}$ is the characteristic recombination time associated with this ionization path. As in the previous rate equations, both the values of $\sigma_5$ and $\tau_{\rm{rec}}$ have been determined to fit the experimental data of \cite{Duchateau:prb:83:075114}. Based on Fig. 6 of this reference, we improved the comparison between experimental and theoretical results by introducing this additional ionization pathway with the parameters specified in Table II. This comparison has been performed up to intensities of $\sim 60$ TW/cm$^2$, but it is expected to hold for higher ones. Figure \ref{Fig1}(b) indeed compares the MPI rate $\sigma_5 I^5$ with the Keldysh rate using Bloch wave functions in crystals for which the electron/hole mass in the valence band is $\sim 3m_e$ \cite{Mezel:jpcm:25:235501}. We can observe the quite good agreement between the Keldysh and MPI rates for KDP, which confirms the validity of Eq. (\ref{eq:mpi_kdp}) at higher intensities. For comparison, the standard Keldysh rate for silica, plotted in dashed line, is in average about 70 times larger in the intensity range $10 \leq I \leq 200$~TW/cm$^2$.

Finally, impact ionization contributes through
\begin{equation}
\partial_t \rho_{\rm{av}}  = \sigma \rho  I/U_i - \rho_{\rm{av}} /\tau_{\rm{rec}}
\end{equation}
to the total electron density in the conduction band:
\begin{equation}
\rho  = \rho_{\rm{MPI}} + \rho_{1} + \rho_{2} + \rho_{\rm{av}} .
\end{equation}
\end{subequations}
Ionization losses, including implicit losses through population of SLG1, follow from adapting the Poynting theorem to these four ionization channels: 
\begin{equation}
\begin{split}
\beta(I)  & = 5 \hbar \omega_0 \sigma_5 I^4 (\rho_{\rm{nt}} + (3+2) \hbar \omega_0 \sigma_3 I^2 \rho_{\rm{SLG1}} \\
 & \quad + (2+2) \hbar \omega_0 \sigma_2 I \rho_{\rm{SLG1}} + \hbar \omega_0 \sigma_1 \rho_{\rm{SLG2}} . 
\end{split}
\end{equation}
Here, the sum $(i+j)$ with $i,j=2,3$ accounts for the total number of photons consumed in filling an SLG state and in transferring electrons from this SLG state to the conduction band.

In the KDP ionization scheme, SLG densities $\rho_{\rm{SLG1}}$ and $\rho_{\rm{SLG2}}$ are limited to the saturation density $\rho_{\rm{sat}}=2\times 10^{17}$ cm$^{-3}$ due to the finite density of defects. Thus, in our numerical implementation we impose the additional constraint $\rho_{\rm{SLG2}}(t) \le \rho_{\rm{sat}}$ when solving Eqs.~(\ref{eq:ionization_kdp}). As far as impact ionization is concerned, we estimate the electron collision time $\tau_c \simeq 100$~fs from the experimental data of Ref.~\cite{Duchateau:prb:83:075114}. Compared to silica, this much larger collision time is probably caused by the energy gap $\ge 1.55$~eV in the KDP conduction band~\cite{Aliabad:ijqc:113:865}, hampering MPI (small $\sigma_5$) as well as avalanche ionization. 

Following \cite{Duchateau:prb:83:075114}, the product of electron capture cross-section $\sigma_c$ and average electron velocity $v$ in Eqs.~(\ref{eq:rho2_kdp}) and (\ref{eq:rhotr_kdp}) depends on the local laser intensity $I$. Direct comparison between the rate equations and experimental data suggest a power-law dependence $\sigma_c v = (\sigma_c v)_{\rm ref} (I/I_{\rm ref})^{-3.3}$, with reference intensity $I_{\rm ref} = 42.32$~TW/cm$^2$ and $(\sigma_c v)_{\rm ref}$ given in Table \ref{table:para}. This behavior can be attributed to the fact that the conduction electrons are strongly heated when operating in the infrared (800~nm). Indeed, the capture cross section depends on the electron kinetic energy, which is function of the laser intensity. Similar temperature-dependent cross-sections can be found measured in \cite{Heppner:pra:13:1000}. However, such a dependency on the laser intensity disappears at ultraviolet wavelengths~\cite{Duchateau:jpcm:25:435501}. Note the recombination times as long as 9 ps used in Table \ref{table:para}: These are attributed to the migration of defects, such as proton migration in the lattice.

\begin{table}[ht]
\begin{tabular}{|r|r|}
\hline Physical parameters & KDP, $U_i$ = 7.7 eV \tabularnewline
\hline $\tau_{\rm{rec}}$ (fs)  & 9000 \tabularnewline
\hline $\sigma_5$ (s$^{-1}$cm$^{10}$W$^{-5}$)  & $1.0 \times 10^{-59}$ \tabularnewline
\hline $\sigma_3$ (s$^{-1}$cm$^{6}$W$^{-3}$)  & $8.6 \times 10^{-27}$ \tabularnewline
\hline $\rho_{\rm{SLG1}}$ (cm$^{-3}$) & $2.0\times 10^{17}$ \tabularnewline
\hline $\tau_{1}$ (fs)  & 300 \tabularnewline
\hline $\sigma_1$ (s$^{-1}$cm$^{2}$W$^{-1}$)  & $2.0$ \tabularnewline
\hline $(\sigma_cv)_{\rm{ref}}$ (s$^{-1}$cm$^{3}$)  & $4.35\times 10^{-7}$ \tabularnewline
\hline $I_{\rm{ref}}$ (TWcm$^{-2}$) & $42.32$ \tabularnewline
\hline $\tau_{\rm{fh}}$ (fs)  & 1000 \tabularnewline
\hline $\sigma_2$ (s$^{-1}$cm$^{4}$W$^{-2}$)  & $1.3 \times 10^{-12}$ \tabularnewline
\hline
\end{tabular}
\caption{Parameters for KDP Eqs.~(\ref{eq:ionization_kdp}) at 800~nm \cite{Duchateau:prb:83:075114}.}\label{table:para}
\end{table} 

\begin{figure}
\includegraphics[width=\columnwidth]{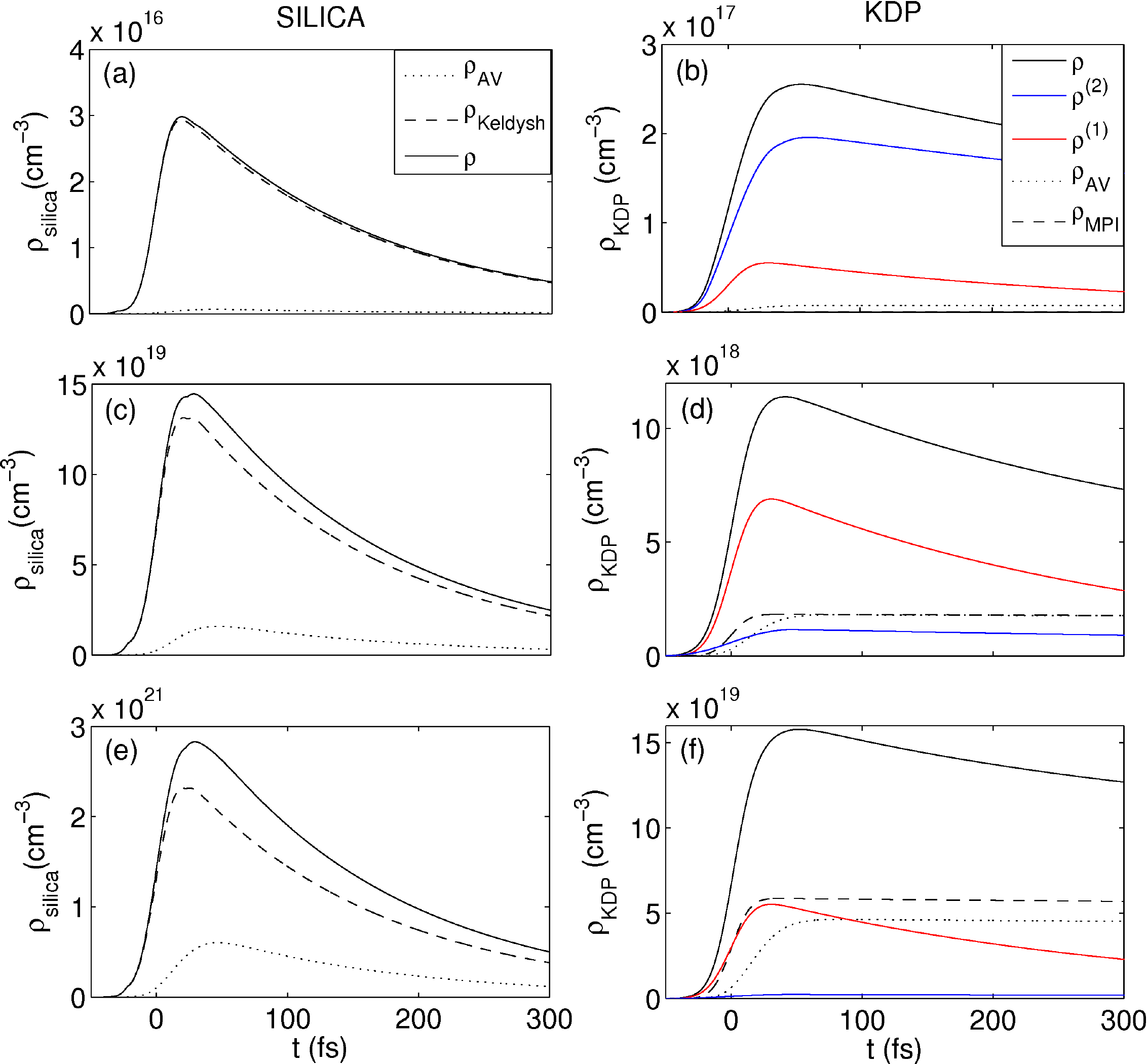}
\caption{Electron densities in the conduction band (blue solid curves) and partial contributions from different transition channels obtained from Eqs.~(\ref{silica}) and (\ref{eq:ionization_kdp}) using Gaussian pulses $I(t) = I_0 \exp(-2t^2/t_p^2)$ with $t_p = 50$~fs and (a,b) $I_0 = 10$~TW/cm$^2$; (c,d) $I_0 = 50$~TW/cm$^2$, and (e,f) $I_0 = 100$~TW/cm$^2$.} 
\label{Fig2}
\end{figure}

Figure~\ref{Fig2} illustrates the plasma response computed from the above models for silica and KDP at intensity levels capable of competing with Kerr self-focusing. For Gaussian intensity profiles $I(t) = I_0 \mbox{e}^{- 2 t^2/t_p^2}$ with 1/e$^2$ half-width duration $t_p = 50$~fs, all partial electron densities contributing to the total conduction band density have been plotted. The choices of $I_0=50$~TW/cm$^2$ and $I_0=100$~TW/cm$^2$ are reasonable clamping intensities for silica and KDP, respectively (see below). We can infer from Fig.~\ref{Fig2} three characteristic intensity regimes. At relatively low intensities up to $10$~TW/cm$^2$ [see Fig.~\ref{Fig2}(a,b)] electron densities in the conduction band of KDP are much higher than in silica thanks to the contributions of the SLGs (see red [light gray] and blue [dark gray] solid lines). However, such density levels are not sufficient to stop Kerr self-focusing. At higher intensities around $I_0=50$~TW/cm$^2$ [see Fig.~\ref{Fig2}(c,d)] we find that direct ionization from the valence band in silica becomes much more efficient, while contributions from the SLGs in KDP saturate due to the limited densities of the defect states. Electron densities in the conduction band of silica are then about 10 times higher than in KDP for this intermediate regime, and we will see below that they are sufficiently high to clamp the laser intensity around $I_0=50$~TW/cm$^2$ in silica. At even higher intensities [see Fig.~\ref{Fig2}(e,f)] direct MPI from the valence band (black dashed curve), even though the cross section $\sigma_5$ is small, becomes the dominant transition process in KDP and elevates the conduction band densities for intensity clamping. We can notice from Fig.~\ref{Fig2} that impact ionization (black dotted lines) contributes always less than 1/3 of the total conduction band density, and plays only a minor role in any of the scenarii described above.

\section{Filamentation of ultrashort (femtosecond) pulses} \label{Sec3}

\begin{figure}
\centerline{\includegraphics[width=\columnwidth]{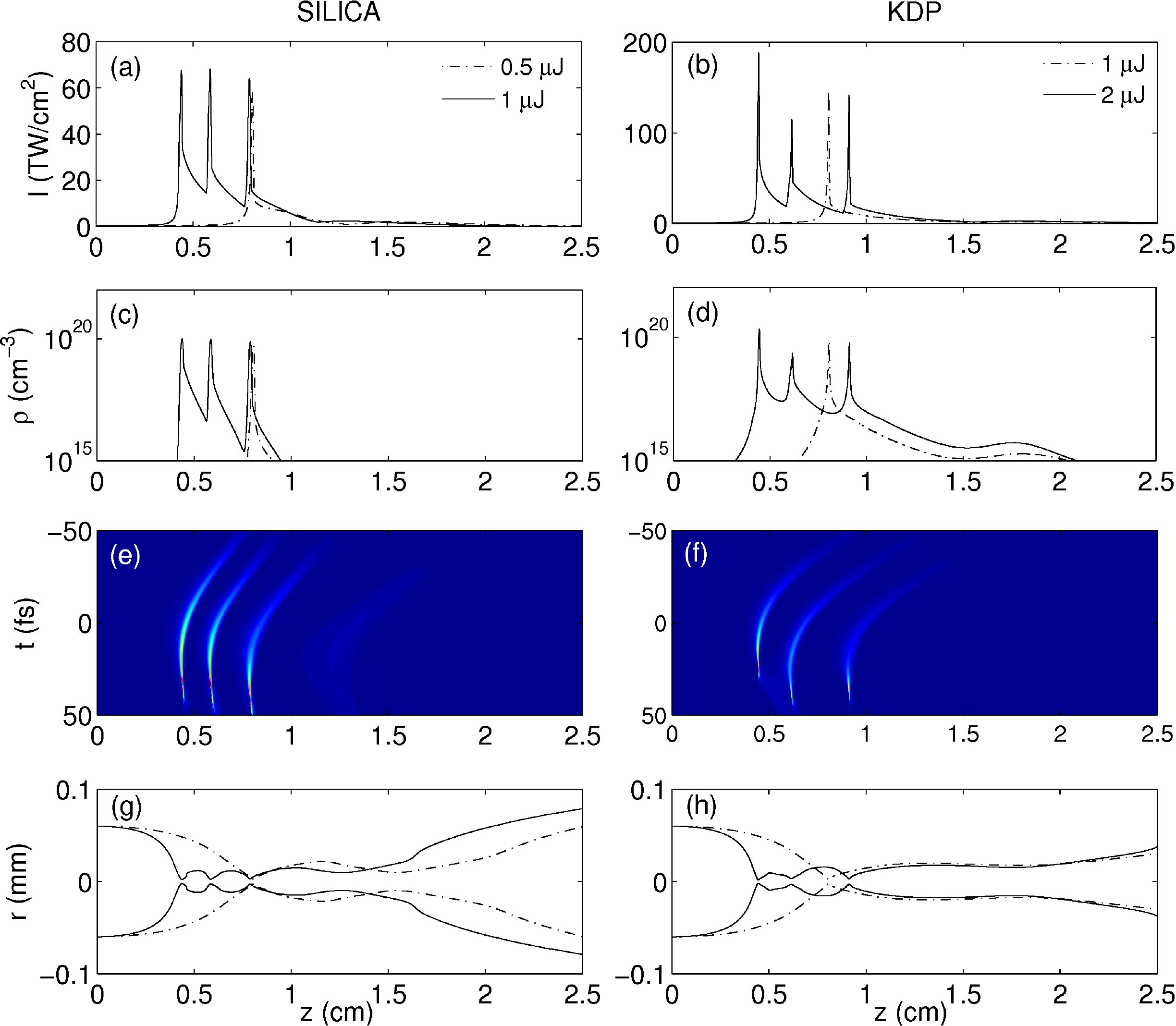}}
\caption{Filamentation dynamics of 50-fs, 60-$\mu$m Gaussian pulses propagating in silica (left column) or KDP (right column). Input pulse energies are $0.5$ $\mu$J (1~$\mu$J) for silica and 1 $\mu$J (resp.\ 2~$\mu$J) for KDP. (a,b) Maximum pulse intensity; (c,d) peak electron density; (e,f) nonlinear on-axis dynamics in the $(t,z)$ plane for the high energy pulses; (g,h) corresponding 1/e$^2$ filament diameters.} 
\label{Fig3}
\end{figure}

We consider Gaussian input pulsed beams with $t_p=50$~fs and $w_0=60~\mu$m. Because the filamentation dynamic intrinsically depends on the ratio of the input peak power over critical, we will compare simulation results for two energy values ensuring similar power ratios. Since fused silica has a nonlinear Kerr index $n_2$ about twice that of KDP, this leads us to double the pulse energy content in KDP simulations.
Figure~\ref{Fig3}(a,b) shows the behavior of the maximum intensities versus propagation distance for pulse energies between $0.5$ and 2~$\mu$J, i.e, with powers up to seven times the critical power. First, we clearly observe that the maximum clamping intensity in KDP is about three times higher than in silica. This directly follows from comparing the ionization cross-sections. Indeed, considering that above 50~TW/cm$^2$ direct 5-photons transitions prevail in KDP, we can make use of the following easy estimate, valid in the MPI regime~\cite{Berge:rpp:70:1633}
\begin{equation}
\label{estimate}
I_{\rm clamp}^{K-1} \approx \frac{2 n_0 n_2 \rho_c}{\Delta \tau \sigma_K \rho_{\rm nt}},
\end{equation}
where $K$ is the photon number, to evaluate the clamping value $I_{\rm clamp}^{\rm KDP} \approx 140$~TW/cm$^2$ in KDP, assuming an efficient plasma defocusing over short times $\Delta \tau \sim 10$~fs~\cite{Ward:prl:90:053901}. This intensity value is higher than that reached in silica, $I_{\rm clamp}^{\rm silica} \approx 60$~TW/cm$^2$, computed from Eq.~(\ref{estimate}) using the formal rescaling $\sigma_6^{\rm silica} \simeq 70 \sigma_5^{\rm KDP}/I_{\rm clamp}$ [see Fig.~\ref{Fig1}(b) and related comments].

Figure~\ref{Fig3} also shows that, for the same power ratio over critical, pulses in silica and KDP develop almost identical propagation dynamics, down to the number of intensity spikes (i.e., focusing-defocusing cycles) and first nonlinear focii located at $z_c \simeq 0.8$~cm at the lowest energy and $z_c \simeq 0.44$~cm at the highest one. Even though similar propagation behaviors can be expected, from simple power scaling arguments and because the Rayleigh length is similar for both materials ($z_R \simeq 2.1$~cm), the quality of these similarities is somewhat surprising and manifests once more the universality and scalability of femtosecond filaments.

The evolution of the conduction band electron density displayed in Figs.~\ref{Fig3}(c,d) is characterized by peak values in KDP being around twice those in silica (e.g., $\rho_{\rm max}^{\rm KDP} = 2\times \rho_{\rm max}^{\rm silica} =2\times 10^{20}$ cm$^{-3}$ at $z \simeq 0.44$ cm for the most energetic pulses). Also, this figure reveals the underlying action of the SLG transition mechanisms: At low intensities $\lesssim 10$~TW/cm$^2$ a postionization regime occurs in KDP after the main MPI range developing at distances $z \leq 1$ cm. Beyond this distance, a residual plasma channel of density $\lesssim 10^{16}$~cm$^{-3}$ is maintained over a propagation length of 1~cm. Pulses in KDP seem able to preserve a longer self-guided state, as is visible from the 1/e$^2$ pulse diameter shown in Figs. \ref{Fig3}(e,f). Moreover, the temporal on-axis dynamics shown in Fig.~\ref{Fig3}(e,f) evidence rapid dispersion of the pulse around each nonlinear focus, which is typical of normal dispersion in transparent solids~\cite{Berge:pre:71:065601}. Strong dispersion takes place over $\sim 0.15$~cm in silica and $\sim 0.175$~cm in KDP ($k^{(2)}_{\rm KDP} < k^{(2)}_{\rm silica}$).
By using these distances as typical values for the dispersion length of self-focused pulses, we can expect occurrence of light structures as short as $\sim \sqrt{0.15 \times k^{(2)}} \approx 7$~fs durations. As far as dissipation is concerned, we report energy losses in KDP higher ($< 30 \%$) than in silica ($< 10\%$), which are probably associated with higher clamping intensities and more ionization channels involved.

\begin{figure}
\centerline{\includegraphics[width=\columnwidth]{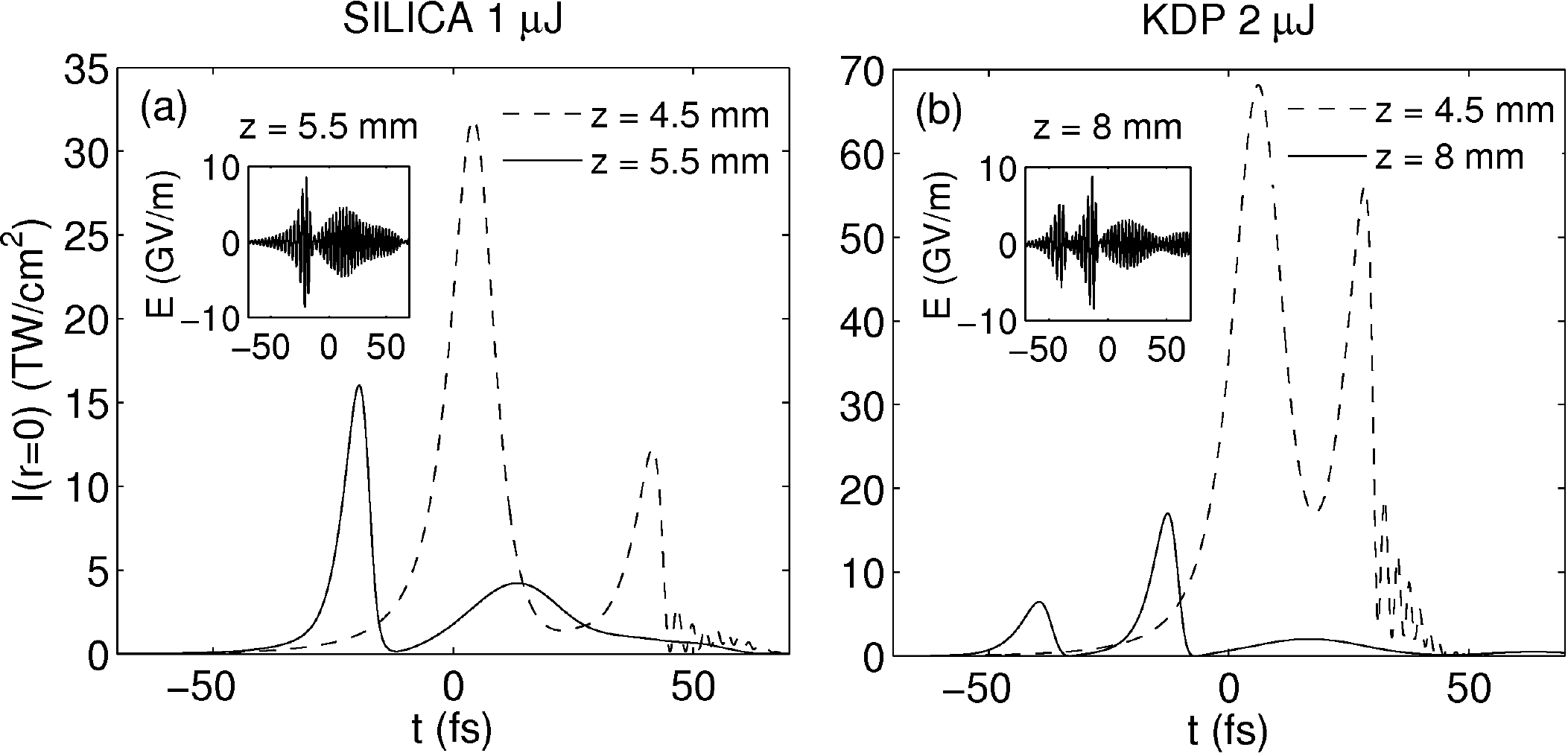}}
\caption{On-axis temporal profiles at two different longitudinal distances for the 50-fs pulse propagating in (a) silica and (b) KDP. Insets show the on-axis electric field of the most compressed pulse.} 
\label{Fig4}
\end{figure}

Figures~\ref{Fig4}(a,b) illustrate two snapshots of on-axis temporal profiles, detailing in particular those with maximum compression in time, i.e., 6.4~fs at $z = 5.5$~mm in silica (1~$\mu$J pulse energy) and 6.2~fs at $z = 8$~mm in KDP (2~$\mu$J pulse energy). Corresponding high-frequency electric fields are presented as insets. We also report an already compressed waveform down to 7.1 fs in KDP at $z = 5.5$ mm (not shown). Basically, the pulse temporal profile follows the well-known dynamic spatial replenishment scenario~\cite{Mlejnek:ol:23:382} relying on the formation of a two-peaked profile after refocusing of the rear pulse, once the latter has been anteriorily defocused by plasma generation. We can notice the shock structures occurring in the trailing edges of the pulses at the shortest distances, which signals a significant action of self-steepening. 

\section{Filamentation of longer (subpicosecond) pulses} \label{Sec4}

As can be seen from Fig.~\ref{Fig2} and Table~\ref{table:para}, ionization dynamics in KDP crystals feature low relaxation times, i.e., free electron lifetimes extend over the picosecond range for the SLG2 defect state transitions and even over almost 10 ps for direct MPI transitions and impact ionization. It is thus instructive to repeat the previous simulations for much longer pulses, in order to see whether the pulse length may interact with the different ionization channels in KDP. 

\begin{figure}
\centerline{\includegraphics[width=\columnwidth]{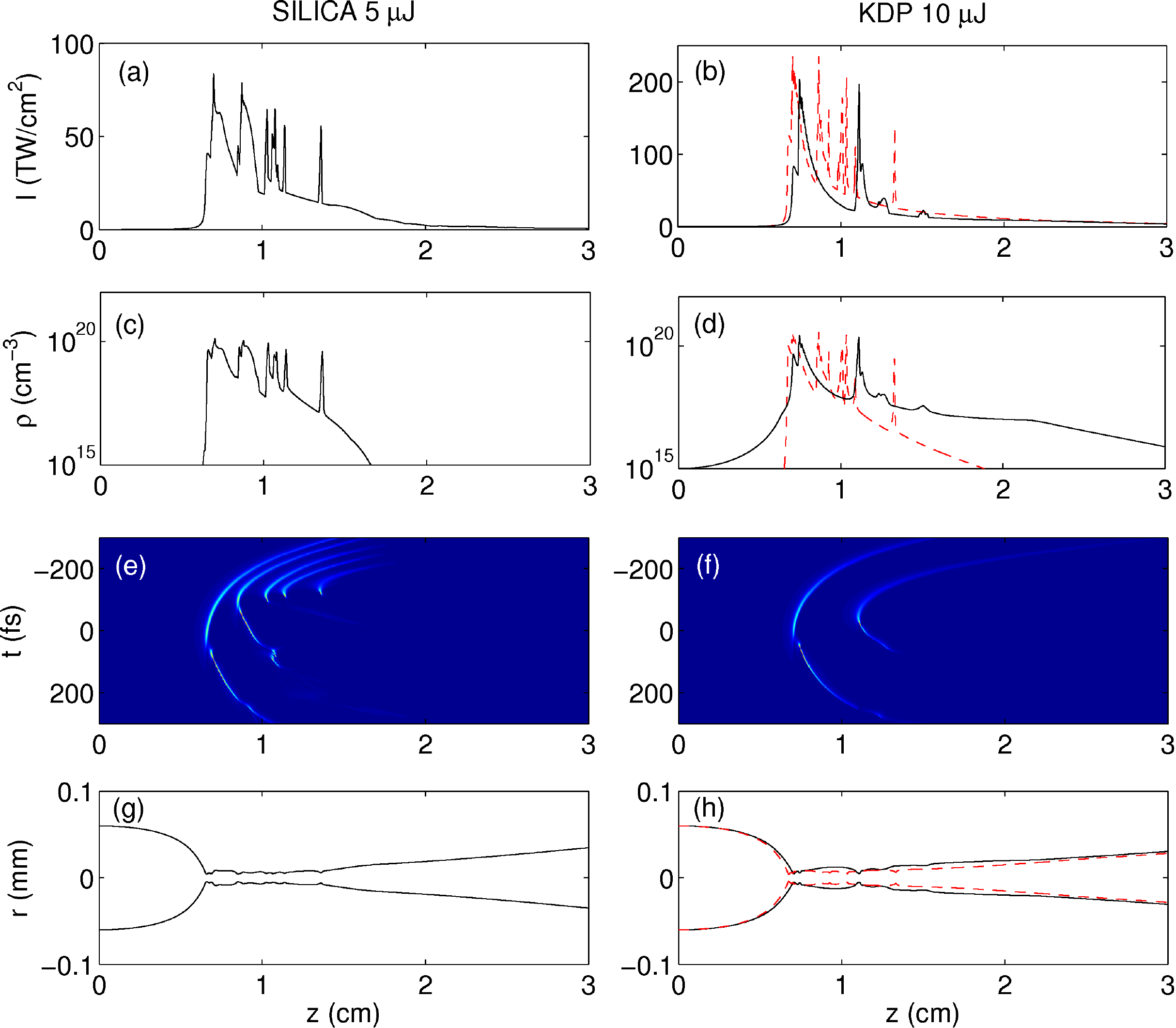}}
\caption{Filamentation dynamics of 500-fs, 60-$\mu$m Gaussian pulses propagating in silica (left column) or KDP (right column). Input pulse energies are $5~\mu$J for silica and 10~$\mu$J for KDP. (a,b) Maximum pulse intensity; (c,d) peak electron density; (e,f) nonlinear on-axis dynamics in the $(t,z)$ plane; (g,h) corresponding 1/e$^2$ filament diameters. In (b,d,h) the red (gray) dashed curve refers to filamentation in KDP with the 5-photon MPI process only.} 
\label{Fig5}
\end{figure}

Figure~\ref{Fig5} illustrates the same pieces of information as Fig.~\ref{Fig3}, but for a 500-fs Gaussian pulse with same initial width and power ratios, i.e., 5~$\mu$J energy in silica and 10~$\mu$J in KDP. Comparable maximum clamping intensities, namely $\sim 80$~TW/cm$^2$ in silica and $\sim 200$~TW/cm$^2$ in KDP are retrieved. This finding is not suprising because the balance between Kerr self-focusing and plasma defocusing is not determined by the initial pulse duration, but by the effective pulse length undergoing ionization at nonlinear focus. Therefore, similar peak density values emerge near the first focus, i.e., $\rho_{\rm max}^{\rm KDP} = 2\times \rho_{\rm max}^{\rm silica} =2.6\times 10^{20}$ cm$^{-3}$ at $z \simeq 0.7$ cm. In contrast, the initial pulse duration affects the number of focusing-defocusing events, dispersion and the filament length. We report up to 15~$\%$ losses in silica and $\sim 40\%$ in KDP over the main filament range, along which multiphoton absorption more severely decreases the pulse power as many time slices participate in the ionization process. From Fig.~\ref{Fig5}(d) we again retrieve an extended plasma channel in KDP ($\rho_{\rm max} \gtrsim 10^{16}$ cm$^{-3}$) that prolongs self-channeling at low intensities $\approx 10$~TW/cm$^2$ [Fig.~\ref{Fig5}(b)], as for the 50-fs pulses. The red dashed curve plots the peak intensity and electron density obtained when only accounting for the 5-photon MPI process in the ionization of KDP. As expected, we can see that the SLG contributions are crucial for supporting this extended plasma range. Figure~\ref{Fig5}(e,f) details the on-axis pulse dynamics for the two materials. Each focusing-defocusing event spans a dispersive sequence extending over slightly longer distances, which does not prevent the pulse from reaching very short durations after several splitting events. Figure \ref{Fig5}(g,h) compares the 1/e$^2$ filament diameter, which here remains of comparable size for the two materials. 

\begin{figure}
\centerline{\includegraphics[width=\columnwidth]{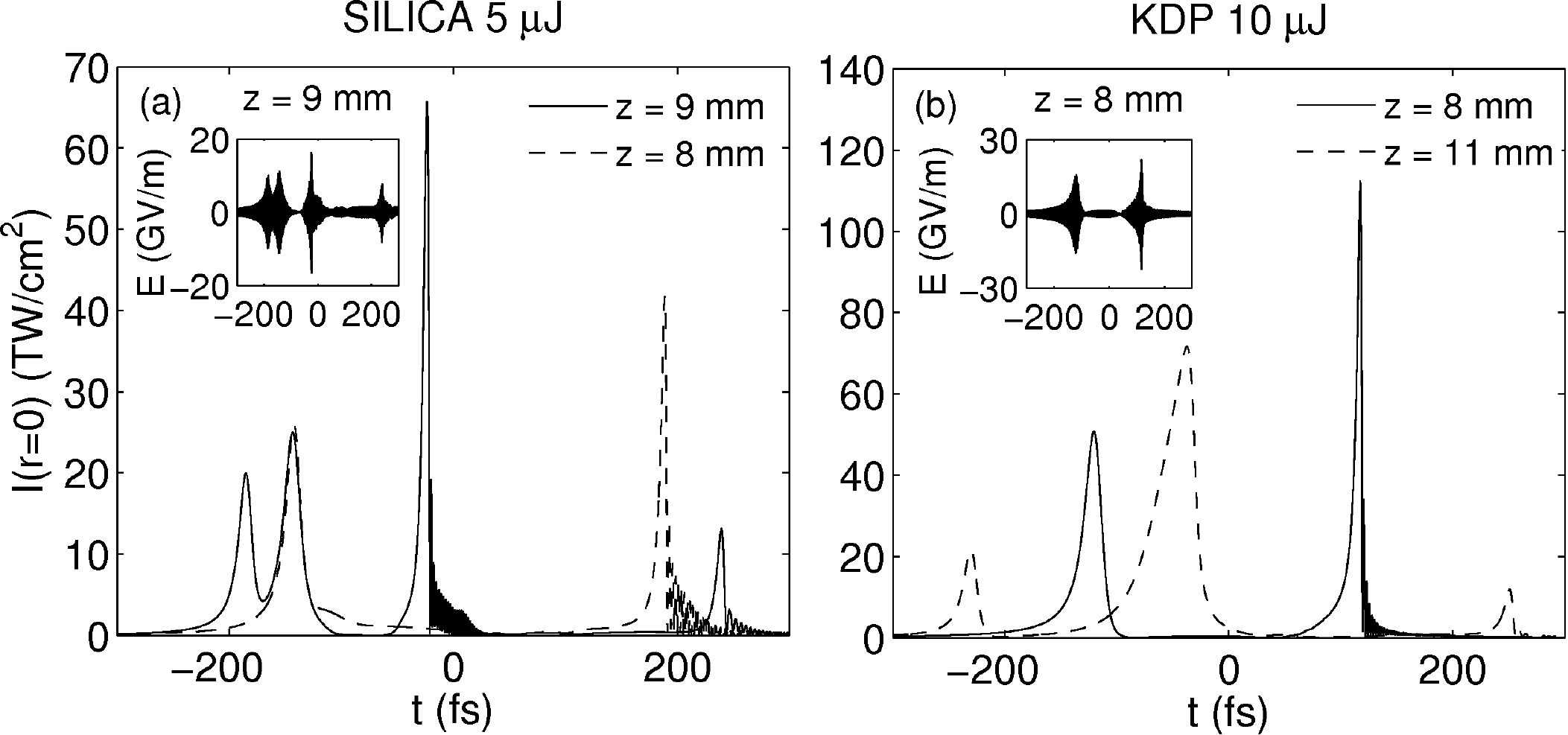}}
\caption{On-axis temporal profiles at two different longitudinal distances for the 500-fs pulse propagating in (a) silica and (b) KDP. Insets show the on-axis electric field of the most compressed pulse.} 
\label{Fig6}
\end{figure}

\begin{figure}
\centerline{\includegraphics[width=\columnwidth]{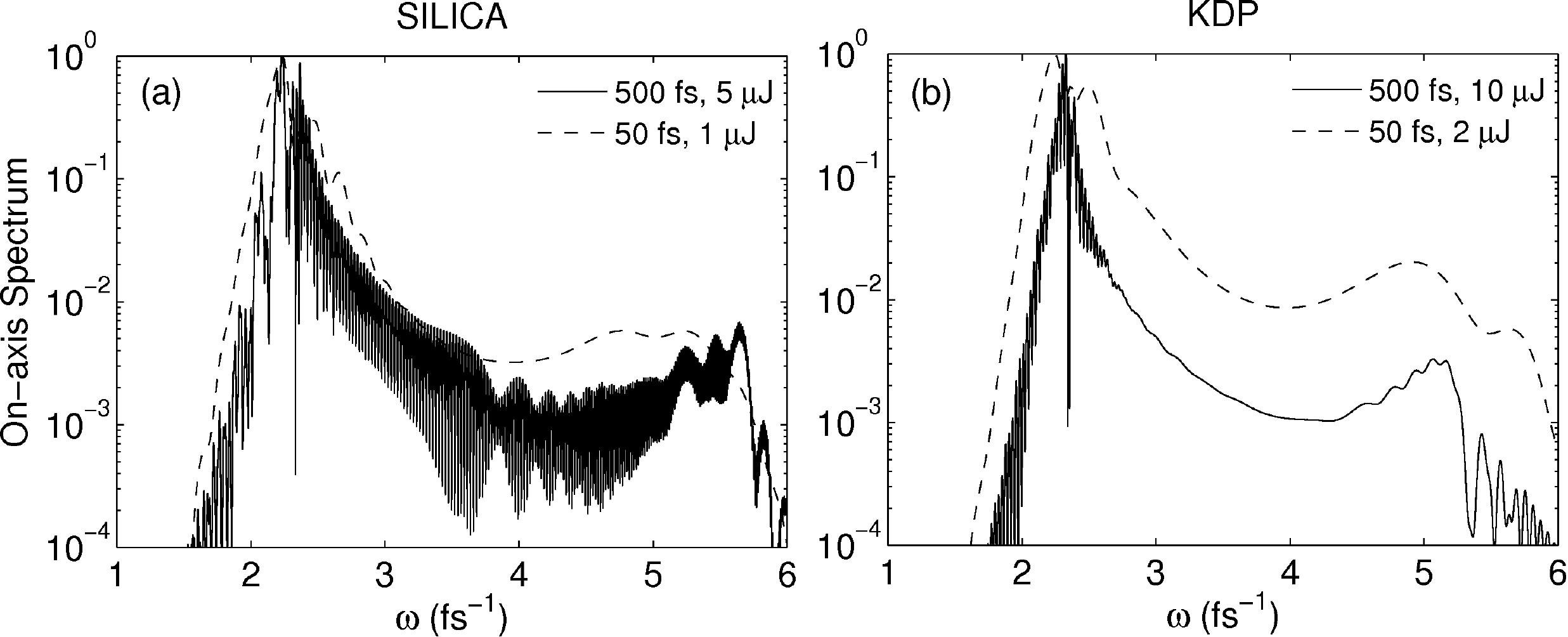}}
\caption{On-axis spectra of the 50-fs and 500-fs long pulses plotted as dashed curves in Fig.~\ref{Fig4} and as solid curves in Fig. \ref{Fig6}. Propagation distances are (a) $z = 4.5$ mm and $z = 9$ mm for silica, and (b) $z = 4.5$ mm and $z = 8$ mm for KDP.} 
\label{Fig7}
\end{figure}

Figures~\ref{Fig6}(a,b) detail the temporal pulse profiles at different propagation distances, including that of maximum compression. Long pulses decay into multi-peaked structures along the optical path, some of which can reach FWHM durations as short as 5.6 fs at $z = 9$ mm in silica and 5.3 fs at $z = 8$ mm in KDP. 
Such self-compression events are expected to produce important supercontinuum generation. Figure~\ref{Fig7}(a,b) presents maximum spectral broadenings reached by the 50-fs and 500-fs pulses in silica and their counterparts in KDP. Close to the peak intensities, strong plasma generation and self-steepening effects blueshift the wings of the spectra~\cite{Rae:pra:46:1084,Anderson:pra:27:1393}. Note the strong blueshifts caused by steep trailing edges~\cite{Skupin:pd:220:14} and oscillations in the spectra, which come from the development of multiple peaks in the temporal profiles~\cite{Tzortzakis:prl:87:213902}.
As a result, the spatio-temporal distortions undergone by our short pulses are mostly generic for longer ones and so are their respective frequency variations. Furthermore, enhancing the input pulse length does not proportionally increase the influence of collisional ionization, as the most intense pulse peaks are supported by very short optical structures $\sim 10$ fs. Despite the fact that collisional time in KDP is 5 times greater than that in silica, we indeed checked that impact ionization does not alter the overall electron density by more than $\sim 8$ $\%$ in silica and by $\sim 2$ $\%$ in KDP near their first nonlinear focus.

\begin{figure}
\centerline{\includegraphics[width=\columnwidth]{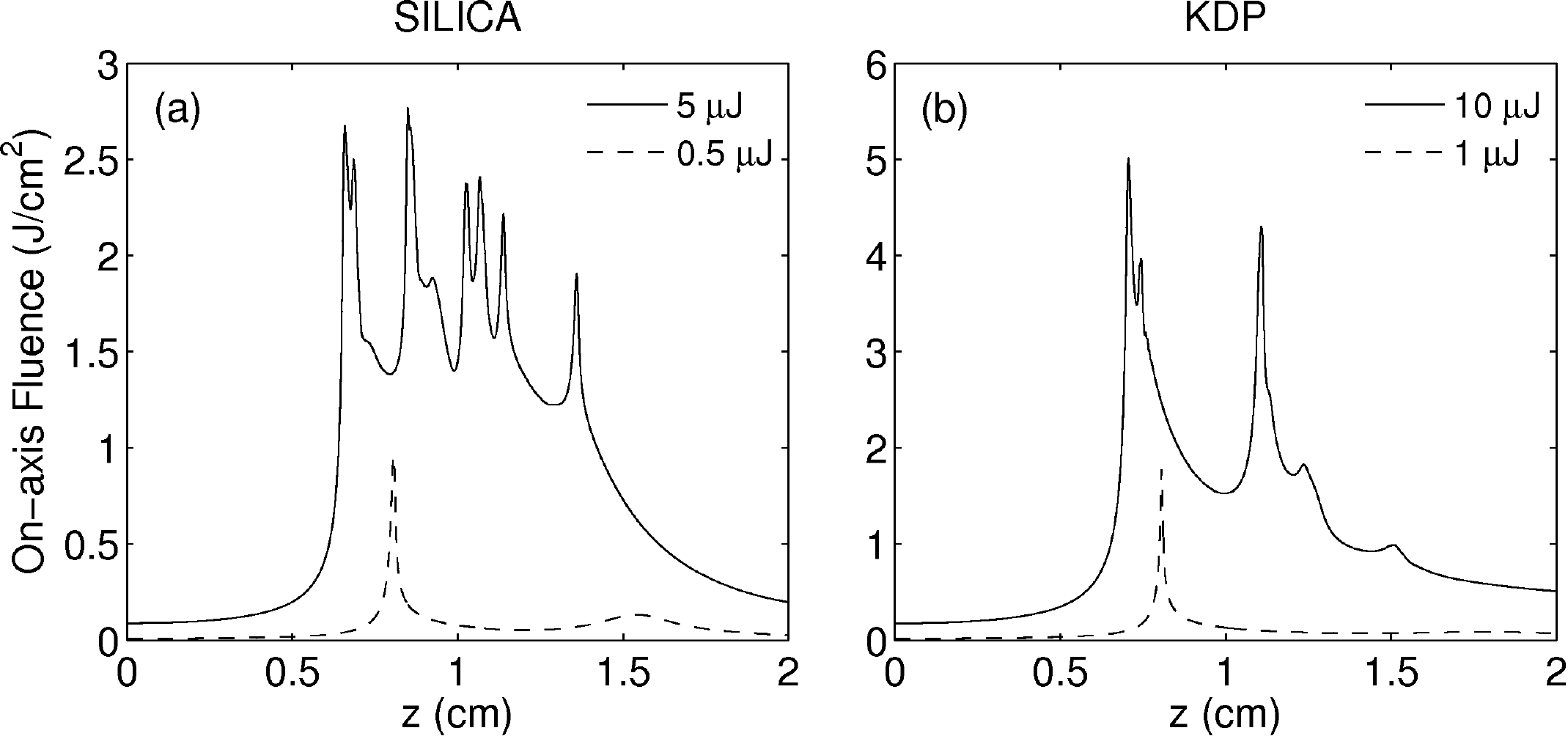}}
\caption{Evolution of the fluence vs. propagation distance for the 50-fs and 500-fs long pulses shown in Fig.~\ref{Fig3} and Fig. \ref{Fig5}.} 
\label{Fig8}
\end{figure}

To end with, we also did not observe significant differences in the fluence evolution [${\cal F} = \int I(r,z,t) dt$] in silica or in KDP (see Fig.~\ref{Fig8}). Typically, 50-fs pulses in KDP reach the maximum fluence level of 2~J/cm$^2$ and about half this value in silica. Using 10 times longer pulses leads to increase the fluence maxima by a factor $\sim 2.5$ only. Based on Fig. 1 of Ref. \cite{Duchateau:jpcm:25:435501}, we expect ablation processes to start at fluences close to 4 J/cm$^2$ with $\sim 50$-fs 800-nm pulses. This damage threshold can only be very locally attained with our long pulses and remain unexceeded with the short ones. The same conclusion applies to fused silica, for which, according to Fig. 9 of Ref. \cite{Penano:pre:72:036412} and related definition of damage thresholds, the fluence threshold is $\sim 1.4$ J/cm$^2$ for 50-fs pulses and about 5 J/cm$^2$ for 10 times longer pulses.

\section{Conclusion} \label{Sec5}

In conclusion, we have investigated numerically the influence of different ionization scenarii in glass (fused silica) and in crystal (KDP) on the filamentation dynamics of femtosecond and subpicosecond pulses. We reported fluence levels remaining below expected damage thresholds in solids. Inside the filaments we observe significantly higher clamping intensities in KDP and higher peak electron densities in the conduction band. This finding may be important with respect to crystal defects or dopants with lower damage thresholds. Despite the increased intensities during propagation, the filamentation dynamics are very similar in both materials if one takes into account the higher critical power in KDP. We provided evidence of the possibility to compress efficiently femtosecond pulses and develop wide supercontinua in both media. Electronic states located in the band gap of KDP increase electron densities in the conduction band at low intensities and thus prolong the plasma channels in a postionization regime. Our results underline the universality of the filamentation characteristics for wide-band gap materials.

This work was performed using High Performance Computing (HPC) resources of TGCC/CCRT. It was granted under the allocation 2013-x2013057027 made by GENCI (Grand Equipement National de Calcul Intensif).

\end{document}